# The effects of chemical pressure on the magnetic ground states of the osmate double perovskites SrCaCoOsO$_6$ and Ca$_2$CoOsO$_6$

Ryan Morrow[1], Jiaqiang Yan[2], Michael A. McGuire[2], John W. Freeland,[3] Daniel Haskel,[3] Patrick M. Woodward[1*]

[1]Department of Chemistry and Biochemistry, The Ohio State University, Columbus, Ohio 43210-1185, USA

[2]Materials Science and Technology Division, Oak Ridge National Laboratory, Oak Ridge, TN 37831, USA

[3]Advanced Photon Source, Argonne National Laboratory, 9700 Cass Avenue, Argonne, Illinois 60439, USA

**Abstract**

The magnetic ground state in the double perovskite system Sr$_{2-x}$Ca$_x$CoOsO$_6$ changes from an antiferromagnet (x = 0), to a spin glass (x = 1), to a ferrimagnet (x = 2) as the Ca-content increases. This crossover is driven by chemical pressure effects that control the relative strength of magnetic exchange interactions. The synthesis, crystal structure, and magnetism of SrCaCoOsO$_6$ and Ca$_2$CoOsO$_6$ are investigated and compared with Sr$_2$CoOsO$_6$. Both compounds adopt a monoclinic crystal structure with rock salt ordering of Co$^{2+}$ and Os$^{6+}$ and $a^-a^-b^+$ octahedral tilting, but the average Co−O−Os bond angle evolves from 158.0(3)° in SrCaCoOsO$_6$ to 150.54(9)° in Ca$_2$CoOsO$_6$ as the smaller Ca$^{2+}$ ion replaces Sr$^{2+}$. While this change may seem minor it has a profound effect on the magnetism, changing the magnetic ground state from antiferromagnetic in Sr$_2$CoOsO$_6$ ($T_{N1}$ = 108 K, $T_{N2}$ = 70 K), to a spin glass in SrCaCoOsO$_6$ ($T_{f1}$ = 32 K, $T_{f2}$ = 13 K), to ferrimagnetic in Ca$_2$CoOsO$_6$ ($T_C$ = 145 K). In the first two compounds the observation of two transitions is consistent with weak coupling between the Co and Os sublattices.

## I. INTRODUCTION

Magnetically frustrated materials have received considerable attention in recent years owing to the manifestation of exotic magnetic states such as spin liquids, spin ices, and spin glasses [1-3]. In particular, the double perovskite structure, $A_2BB'O_6$, which consists of rock salt ordered $BO_6$ and $B'O_6$ corner connected octahedra [4] has been a rich vein of magnetically frustrated materials. When the B cation is diamagnetic the resulting B′ sublattice possesses a geometrically frustrated fcc topology. The competition between nearest-neighbor and next-nearest-neighbor exchanges can lead to various patterns of antiferromagnetic (AFM) order [5] as well as more exotic magnetic ground states, such as a spin-singlet state (i.e. $Ba_2YMoO_6$), incommensurate magnetic order (i.e. $La_2NaOsO_6$), and an unconventional spin glass (i.e. $Sr_2MgReO_6$) [6-8]. When both B and B′ are magnetic transition metal ions the resulting B−O−B′ superexchange pathway must also be considered. When the B−O−B′ superexchange is strong, as is the case for $Sr_2CrOsO_6$ ($T_C$ = 725 K), high temperature ferrimagnetism is observed [9]. When it is very weak, as is the case for $Sr_2CoOsO_6$, AFM ordering of weakly coupled B and B′ sublattices results [10, 11]. When the B−O−B′ interaction is intermediate between these two extremes, highly frustrated magnetic ground states and incommensurate magnetic order are expected to emerge.

Recently, it has been shown that the insulator $Sr_2FeOsO_6$ undergoes two antiferromagnetic transitions resulting from a complex interplay of competing exchange interactions [12, 13]. Interestingly $Ca_2FeOsO_6$, which contains the same $Fe^{3+}$ and $Os^{5+}$ ions, has a ferrimagnetic ground state [14, 15]. The change in magnetic ground state is thought to occur because the longer range Fe−O−Os−O−Fe and Os−O−Fe−O−Os interactions become weaker as the Fe−O−Os bond angles become increasingly bent, which allows the antiparallel Fe−O−Os superexchange interaction to gain the upper hand [15]. In this study we explore the effect of chemical pressure on the magnetism of $Sr_2CoOsO_6$, where long range Co−O−Os−O−Co and Os−O−Co−O−Os interactions have been shown to drive magnetic ordering, while nearest neighbor Co−O−Os interactions are particularly weak [10, 11]. Our results, summarized in Figure 1, show that through the application of chemical pressure via substitution of Ca for Sr, the structure is distorted impacting the relative strengths of the superexchange interactions, which in turn converts the ground state from antiferromagnetic to ferrimagnetic through a spin glass region where exchange competition is strongest.

## II. EXPERIMENTAL

Powder samples of approximately 1.6 g of $SrCaCoOsO_6$ and $Ca_2CoOsO_6$ were synthesized by heating the reagents at 1000 °C for a period of 48 hours in sealed silica tubes. $PbO_2$ was included separately as an in-situ oxygen gas source as described in an earlier study [10]. Further synthetic details and precautions are available as Supplemental Material. Phase purity of these products was established using laboratory X-ray diffraction. The high scattering contrast of Co and Os to X-rays allowed us to confirm complete rock salt ordering of Co and Os, while Sr and Ca are disordered on the A site in SrCaCoOsO$_6$. Neutron powder diffraction (NPD) measurements at 10 and 300 K for both materials were conducted on the POWGEN [16] beamline at the Spallation Neutron Source (SNS). Additional temperature dependent neutron powder diffraction data were collected at 15 K intervals from 10 to 205 K on Ca$_2$CoOsO$_6$. Rietveld refinements were performed using the GSAS EXPGUI package [17, 18].

Dc magnetization measurements as a function of temperature were conducted utilizing a Quantum Design Magnetic Property Measurement System (MPMS) SQUID magnetometer under zero field cooled and field cooled (1 kOe) conditions with a measuring field of 1 kOe for both materials. Dc magnetization measurements as a function of field were collected in a field range of ±50 kOe at 5 K for Ca$_2$CoOsO$_6$. Similar data were collected for SrCaCoOsO$_6$ in a field range of ±70 kOe at 2, 20, and 50 K. Heat capacity measurements were collected on a sintered pellet mounted with Apiezon grease using a Quantum Design Physical Property Measurement System (PPMS) using the relaxation technique. Ac magnetization studies at a variety of frequencies were conducted in the temperature range 3 to 60 K using a Quantum Design PPMS. Dc magnetization relaxation measurements were conducted by cooling in an applied field of 10 kOe to specific temperatures before removing the field and collecting data as a function of time after field removal.

### III. RESULTS AND DISCUSSION

Both materials crystallize with monoclinic $P2_1/n$ symmetry resulting from complete ordering of Co$^{2+}$ and Os$^{6+}$ ions and $a^-a^-c^+$ octahedral tilting [19], as defined using Glazer notation [20]. Refined room temperature neutron powder diffraction patterns and depictions of the monoclinic crystal structure are given in Figure 2. Select refined parameters and structural details at 10 and 300 K are listed in Table 1. Further refined parameters, including variable temperature results, are available in the Supplemental Material. No evidence for a structural phase transition was observed in either material over the temperature range examined. The average bond lengths for both CoO$_6$ and OsO$_6$ octahedra are comparable to those found in Sr$_2$CoOsO$_6$ [10], as well as other double perovskites oxides containing Os$^{6+}$ or high spin Co$^{2+}$ [21, 22], confirming that the electronic configuration of these cations does not change significantly upon substitution of Ca for Sr. The low temperature Co−O−Os bond angles, which are dictated by the magnitude of octahedral tilting, take on average values of 157.2° in SrCaCoOsO$_6$ and 149.7° in Ca$_2$CoOsO$_6$. By comparison the Co−O−Os bond angles in Sr$_2$CoOsO$_6$ are 180° in the *c*-direction and 162.8° in the *ab*-plane [10].

Figure 3a shows the temperature dependence of the dc magnetization of $Ca_2CoOsO_6$ as measured in a SQUID magnetometer in an applied field of 1 kOe. $Ca_2CoOsO_6$ exhibits a sharp rise in magnetization with a critical temperature of approximately 145 K. The temperature range above this transition was insufficient for a reasonable Curie-Weiss fit. The field dependence up to 50 kOe at 5 K produces a hysteresis loop with an $M_{sat}$ of 1.77 $\mu_B$ per formula unit (see Figure 3b). For the electronic configurations of high spin $d^7$ for $Co^{2+}$ (S = 3/2) and $d^2$ for $Os^{6+}$ (S = 1) a ferromagnetic $M_{sat}$ should be in the vicinity of 5 $\mu_B$ per formula unit, assuming spin-only moments. On the other hand, an $M_{sat}$ of only 1 $\mu_B$ per formula unit is expected for a ferrimagnetic ground state. The observation of an experimental $M_{sat}$ larger than the expected spin only value can be attributed to the presence of spin-orbit coupling that increases the $Co^{2+}$ moment and decreases the $Os^{6+}$ moment. As discussed later, non-collinear magnetic configurations which might also lead to an increase in $M_{sat}$ are ruled out by the absence of new Bragg peaks in neutron powder diffraction patterns collected below the Curie temperature.

X-ray magnetic circular dichroism (XMCD) measurements were conducted on APS beamlines 4-ID-C and 4-ID-D at the Co and Os $L_{3,2}$ edges at temperatures of 5 and 10 K respectively under fields of ±35 kOe. The relative sign of the XMCD signal, shown as in Figure 4, is reversed between Co and Os, indicating that the moments localized on these elements are oriented antiparallel to one another. Application of the sum rules yields a negative moment on Os and a positive moment on Co, meaning that upon application of a field, the Co moment aligns parallel to the field direction while Os moments align antiparallel to the external magnetic field. This is consistent with a ferrimagnetic configuration, as suggested by the previously described magnetization data. Neglecting $T_z$, the magnetic dipole operator, the calculated ratio of the orbital moment to the spin moment, $m_l/m_s$, is +44% and −28.5% for Co and Os respectively, consistent with expectations for positive and negative orbital contributions for the high spin $d^7$ and $d^2$ electronic configurations. The spin and orbital moment contributions on Os calculated from the sum rules are 1.97(17) and −0.56(4) $\mu_B$ respectively resulting in a substantial Os moment in sharp contrast to the results from NPD refinements. The spin and orbital moments found in this way on Co are 1.8 $\mu_B$ and 0.8 $\mu_B$ respectively. Additionally, the XAS spectrum of the Co $L_{3,2}$ edges is consistent with the assignment of high spin $Co^{2+}$ [23].

Neutron diffraction measurements on $Ca_2CoOsO_6$ taken at intervals of 15 K up to 205 K confirm the magnetic ordering. Plotted against the right axis of Figure 3a is the integrated intensity of the ($\bar{1}01$), (101), and (011) reflections, which contain the strongest magnetic contribution, rising as the sample is cooled below the Curie temperature. Figure 5 shows these peaks at 10 and 300 K for comparison to highlight the increased intensity from the magnetic contribution. No additional purely magnetic reflections were detected at low temperatures, excluding more complicated canted magnetic structure models. Magnetic refinements were conducted using the magnetic form factor coefficients for $Os^{6+}$ as given by Kobayashi *et al*. [24]. Due to the high degree of correlation between the Co and Os moments, it was not possible to meaningfully refine the Co and Os moments independently. If one refines the data with only ferromagnetically aligned spins on cobalt (see Figure 5), a moment of 2.33(8) $\mu_B$ oriented parallel to the b-axis is obtained. While this value is not unreasonable for a high spin $Co^{2+}$ ion, the lack of a moment on Os is inconsistent with the XMCD data and the total moment is not in agreement with the measured value of $M_{sat}$ (1.77 $\mu_B$ per formula unit).

To probe the sensitivity of the NPD data to different combinations of Co and Os moments a series of refinements were performed using fixed Os moments. The results show that fits of comparable quality can be obtained with relatively small Os moments aligned either parallel or antiparellel to a more substantial Co moment (see Supplemental Material Figure S4). Discarding solutions where the Co and Os moments are aligned parallel to each other, as well as those where the Co moment falls below 2 $\mu_B$ suggests that the Os moment lies somewhere between 0 and 0.4 $\mu_B$. The reasons why Os moment estimated from the neutron refinements is significantly smaller than the moment estimated from XMCD is not clear.

The dc magnetization of $SrCaCoOsO_6$ reveals two cusps, each with a field cooled-zero field cooled divergence, at 10 and 27 K (see Figure 6a). The Curie-Weiss fit between 200 and 350 K resulted in an effective moment of 5.51 $\mu_B$ and Weiss constant of −146 K. The effective moment is somewhat higher than the calculated spin only value of 4.80 $\mu_B$, however in light of the XMCD measurements on $Ca_2CoOsO_6$ this increase in effective moment is likely due to a positive orbital contribution that increases the cobalt moment more than the negative orbital contribution decreases the osmium moment [21]. The Weiss constant is very similar to the $T_C$ of $Ca_2CoOsO_6$. No discernible anomalies are apparent in the heat capacity with or without an applied field of 80 kOe, as measured by a PPMS and shown in Figure 6b, nor is any magnetic scattering due to long range ordering apparent in the 10 K neutron powder diffraction pattern shown in Figure 6c.

As the dc magnetization and heat capacity results are indicative of spin glass freezing, ac magnetization measurements were performed as a function of temperature and frequency with an applied DC field of 0 Oe and AC field of 10 Oe. The resulting real and imaginary components, displayed in Figure 7a and 7b, respectively, show a strong frequency dependence at the lower temperature transition, $T_{f1}$=13 K, as well as a weaker frequency dependence at the higher temperature transition, $T_{f2}$=32 K. The frequency and temperature dependent behaviors observed here support the identification of both peaks observed in the dc magnetization measurements as classical spin-glass freezing events [25, 26]. Activation energies ($E_A$) associated with these transitions were determined by modeling the relationship between temperature at which the real part of the magnetization $M'$ peaks ($T_{peak}$) and the frequency (f) using $f(T_{peak})=f_0 \cdot \exp(-E_A/k_B T)$, where $k_B$ is the Boltzmann constant. The results are shown in Figure 8a. An order of magnitude larger activation energy is observed for the higher temperature transition, indicating that the glassy state existing between $T_{f1}$ and $T_{f2}$ is much more "rigid" than that formed upon cooling below $T_{f1}$, which is also consistent with the significantly smaller dissipation observed at $T_{f2}$ as measured by the imaginary part of the magnetization M″ (Figure 7b). The frequency dependence and the magnitude of $M''$ at $T_{f1}$ is sufficiently strong to allow an examination of how the characteristic relaxation time varies with frequency [27], as shown in Figure 8b. This plot shows a single, characteristic peak in frequency which varies smoothly with temperature around $T_{f1}$.

Since spin-glass behavior is confirmed by the ac magnetic measurement, the time dependence of the dc magnetization was examined. Magnetic moment measurements were made after cooling from 50 K to the measurement temperature in a magnetic field of 10 kOe and then reducing the field to zero. Results shown in Figure 9, were fit with a stretched exponential of the form $M \sim \exp[-(t/\tau)^b]$, where $\tau$ is the time constant associated with the relaxation of the magnetic moment [25, 26]. In general, it is expected that the time constant will increase upon cooling deeper into the spin glass state. The temperature dependence of the fitted time constants is shown in the inset of Figure 9. A sharp decrease in τ is observed upon cooling below 10 K, indicating the onset of a second glassy state with faster relaxation rates, consistent with the ac results discussed above. Thus, the magnetic data indicate the formation of two independent spin glass states in SrCaCoOsO$_6$. In the context of what is known about the properties of Sr$_2$CoOsO$_6$ [10], one can speculate that they correspond to independent freezing of the Co and Os sublattices.

## IV. CONCLUSIONS

As was the case for the Sr$_2$FeOsO$_6$ system [14, 15], the competing magnetic interactions are extremely sensitive to changes the Co−O−Os bond angles in response to substitution of Ca for Sr. As the chemical pressure is increased and the bond angles are bent the Co−O−Os−O−Co and Os−O−Co−O−Os exchange pathway strengths are weakened in favor of Co−O−Os superexchange. In the fully Ca substituted end member, Ca$_2$CoOsO$_6$, the competition between the aforementioned long range bond exchange pathways and the shorter AFM Co−O−Os exchange is dominated by the latter resulting in ferrimagnetic order at the moderate $T_C$ of 145 K. In SrCaCoOsO$_6$ where the octahedral tilting is intermediate, the relative strengths of these exchanges are comparable and spin glass states emerge. Interestingly, it appears that the Co−O−Os superexchange pathway remains weak enough to allow the Co and Os sublattices to freeze independently of one another. It is remarkable that these subtle changes in bond angle can produce such dramatic changes in the magnetic ground state of these and related mixed 3d/5d transition metal oxides. Further theoretical and experimental study of these systems is likely to yield additional unusual magnetic phenomena and is needed to increase our understanding of the competing superexchange interactions that are responsible for this rich magnetic phase space.

**Associated Content:**

**Supporting Material**

    Further synthetic details, along with data and analysis of the $Ca_2CoOsO_6$ neutron powder diffraction patterns.

**Author Information:**

    **Corresponding Author**
    woodward@chemistry.ohio-state.edu
    **Notes**
    The authors declare no competing financial interest.

**Acknowledgements:**

Support for this research was provided by the Center for Emergent Materials an NSF Materials Research Science and Engineering Center (DMR-0820414), and the U.S. Department of Energy, Office of Science, Basic Energy Sciences, Materials Sciences and Engineering Division (ac magnetization measurements and analysis). A portion of this research was carried out at Oak Ridge National Laboratory's Spallation Neutron Source, which is sponsored by the U.S. Department of Energy, Office of Basic Energy Sciences. Use of the Advanced Photon Source was supported by the U. S. Department of Energy, Office of Science, Office of Basic Energy Sciences, under Contract No. DE-AC02-06CH11357.

**Figures and Tables**

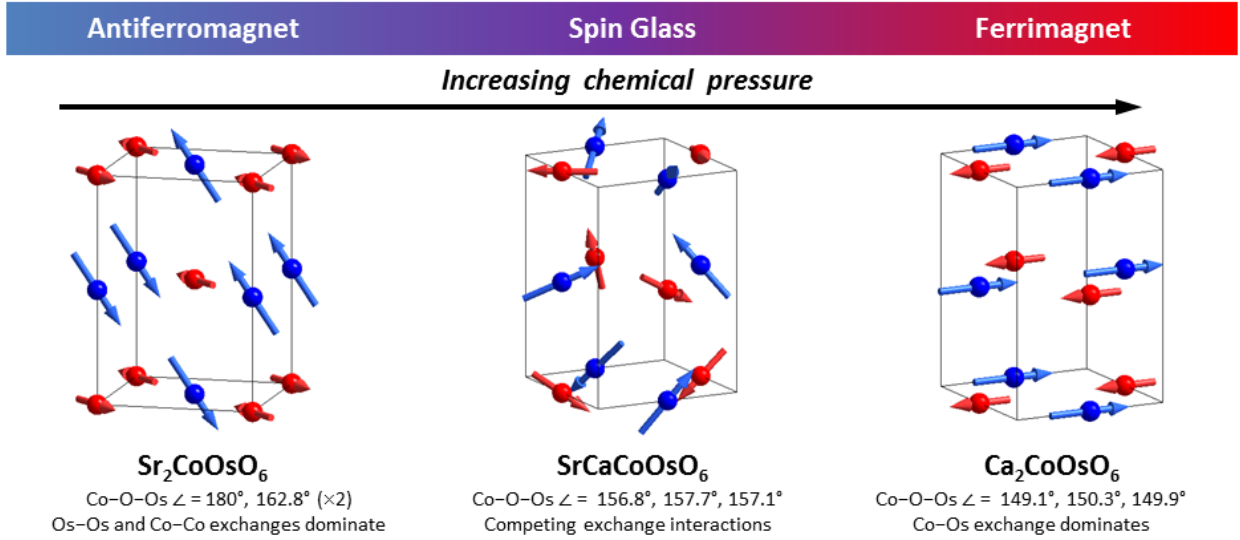

FIG. 1 (color online). A schematic of the evolution of the magnetic ground state of $A_2CoOsO_6$ with chemical pressure due to the reduction of the Co−O−Os bond angle that accompanies increased octahedral tilting. Red and blue spheres correspond to Os and Co, respectively.

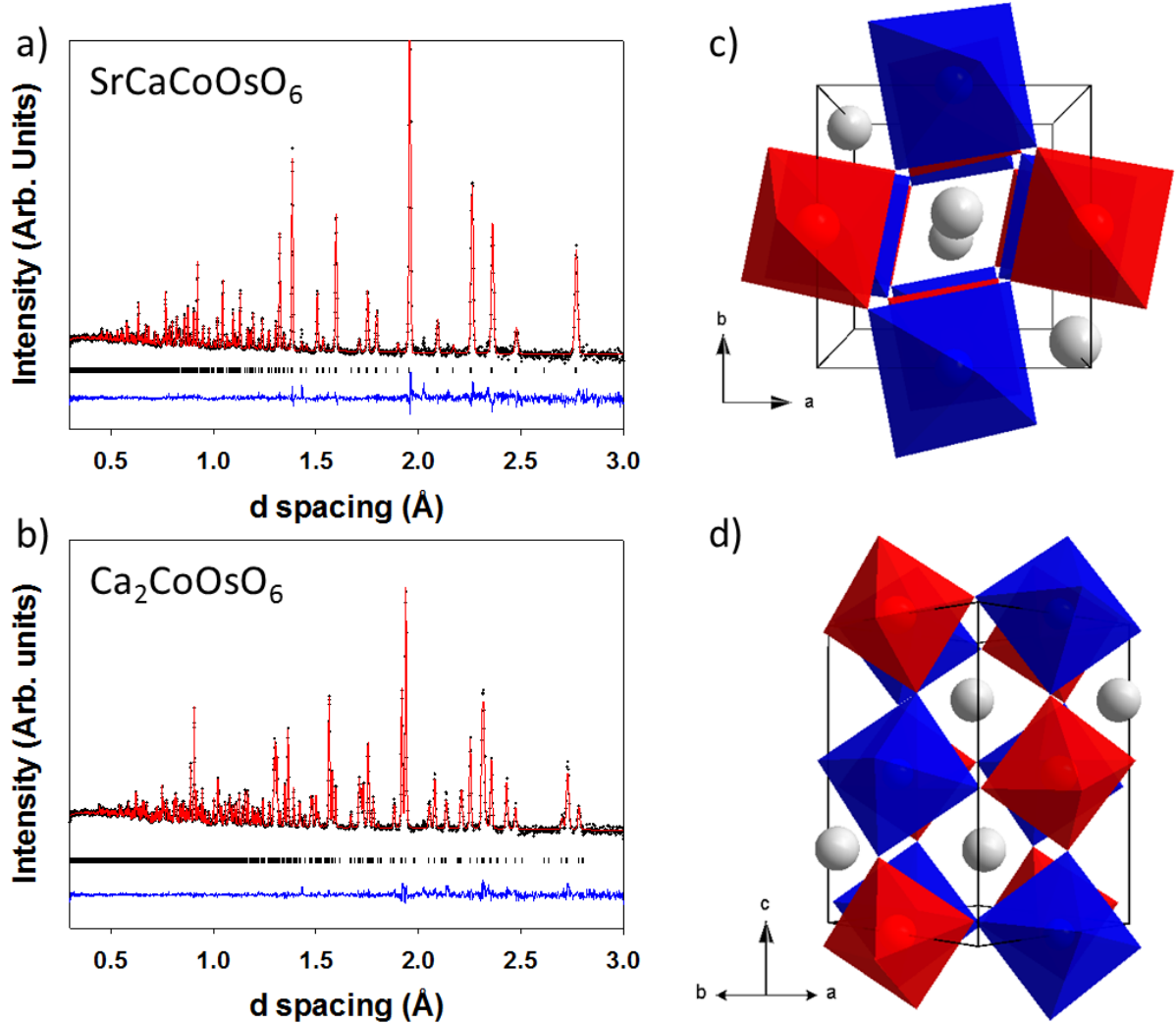

FIG. 2 (color online). Refined room temperature neutron powder diffraction patterns of (a) SrCaCoOsO$_6$ and (b) Ca$_2$CoOsO$_6$. The crystal structure of Ca$_2$CoOsO$_6$ is shown looking down the (c) 001 and (d) 110 directions to illustrate the $a^-a^-b^+$ pattern of octahedral tilting that leads to $P2_1/n$ symmetry.

|  | Ca$_2$CoOsO$_6$ | | SrCaCoOsO$_6$ | |
|---|---|---|---|---|
|  | 10 K | 300 K | 10 K | 300 K |
| Space Group | $P2_1/n$ | $P2_1/n$ | $P2_1/n$ | $P2_1/n$ |
| $a$ (Å) | 5.3753(1) | 5.39236(9) | 5.5211(2) | 5.5219(3) |
| $b$ (Å) | 5.5683(1) | 5.55632(9) | 5.5393(2) | 5.5410(2) |
| $c$ (Å) | 7.6362(2) | 7.6632(13) | 7.7942(3) | 7.8125(3) |
| $\beta$ | 89.755(1) | 89.831(1) | 89.926(6) | 89.917(5) |
| $R_{wp}$ (%) | 2.90 | 2.08 | 2.97 | 3.06 |
| Co−O1 (×2) | 2.080(2) | 2.076(1) | 2.069(3) | 2.061(4) |
| Co−O2 (×2) | 2.072(2) | 2.063(2) | 2.053(3) | 2.051(4) |
| Co−O3 (×2) | 2.032(2) | 2.031(2) | 2.049(4) | 2.063(5) |
| Os−O1 (×2) | 1.935(1) | 1.934(2) | 1.919(3) | 1.926(3) |
| Os−O2 (×2) | 1.932(2) | 1.936(2) | 1.929(3) | 1.927(5) |
| Os−O3 (×2) | 1.922(2) | 1.929(2) | 1.928(4) | 1.921(4) |
| ∠Co−O1−Os | 149.07(9) | 149.86(9) | 156.8(3) | 157.7(3) |
| ∠Co−O2−Os | 150.29(9) | 150.97(9) | 157.7(3) | 158.9(3) |
| ∠Co−O3−Os | 149.86(8) | 150.78(8) | 157.1(2) | 157.4(3) |

TABLE 1. Select structural features as determined from Rietveld refinements of neutron powder diffraction data. The ∠ symbol represents a bond angle.

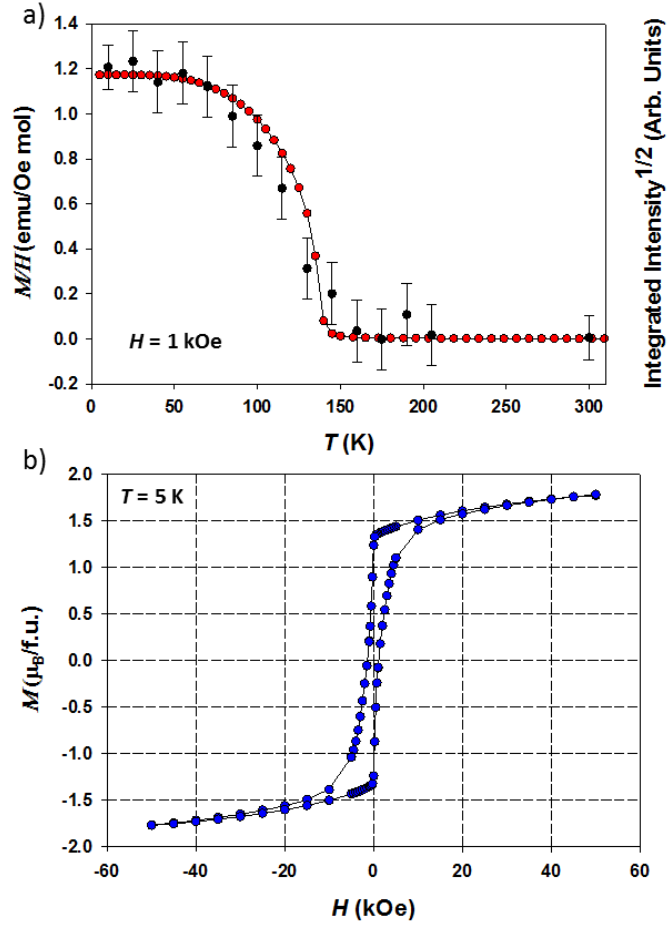

FIG. 3 (color online). (a) The field cooled temperature dependence of the dc magnetization of $Ca_2CoOsO_6$ measured in 1 kOe (red circles, left axis) and the square root of the total integrated intensity of the ($\bar{1}01$), (110), and (011) reflections which increase due to magnetic scattering below the ordering temperature (black circles, right axis) (b) The field dependent magnetization of $Ca_2CoOsO_6$ measured at 5 K.

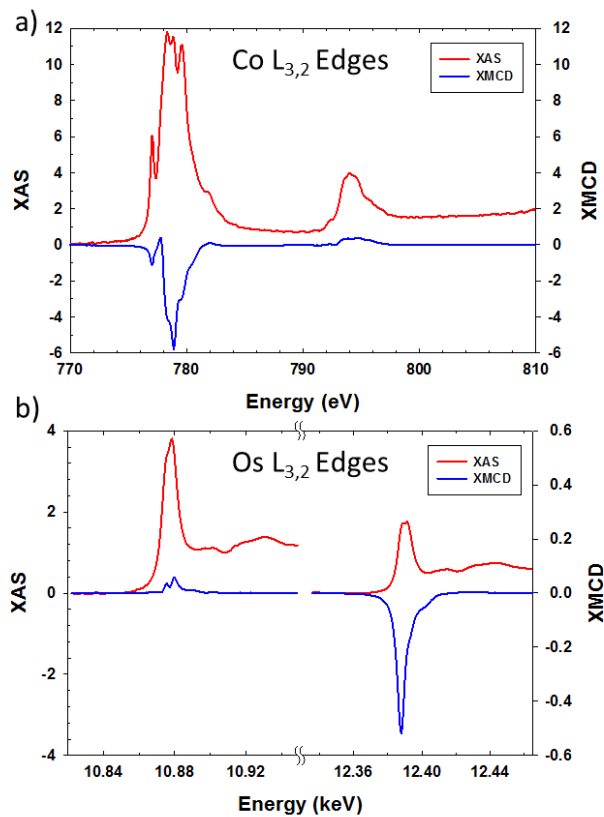

FIG. 4 (color online). XAS and XMCD data for $Ca_2CoOsO_6$ measured at the (a) Co and (b) Os $L_{3,2}$ edges.

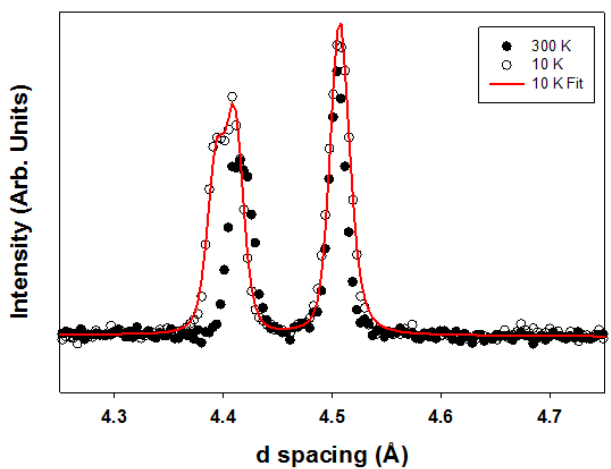

FIG. 5 (color online). High d spacing reflections of $Ca_2CoOsO_6$ at 300 K (filled circles) and 10 K (open circles) showing the increased intensity due to magnetic scattering with fit to 10 K data shown as the red curve. The larger increase in intensity of the 101 as compared to the 011 peak is due to the preferred orientation of the moments along the b direction.

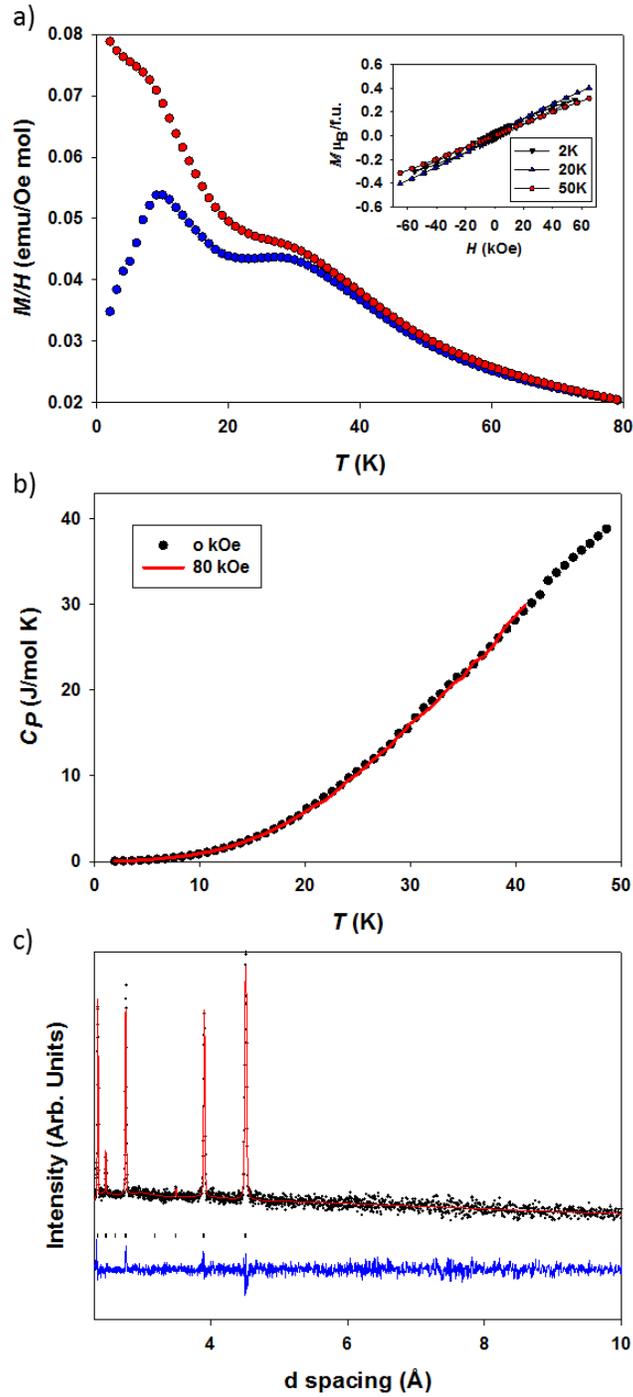

FIG. 6 (color online). (a) The field cooled (red) and zero field cooled (blue) temperature dependence of the dc magnetization of SrCaCoOsO$_6$ measured in 1 kOe, the field dependence at 2, 20, and 50 K is shown in the inset, (b) the heat capacity in 0 (black symbols) and 80 kOe (red line) fields, and (c) the high d-spacing region of the neutron powder diffraction pattern at 10 K (black dots). The agreement of the fit (red curve) to the data obtained using a nuclear only model, signals a lack of long range magnetic order.

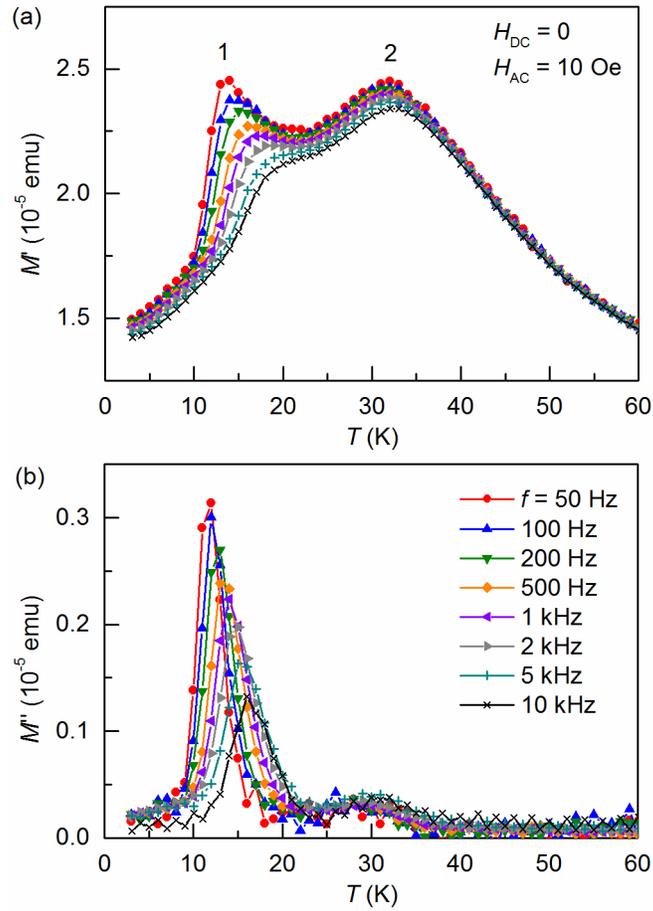

FIG. 7 (color online). Results from frequency and time dependent magnetization measurements on SrCaCoOsO$_6$. The real (M') and imaginary (M'') parts of the ac magnetization are shown in (a) and (b), respectively. Measurements were performed in zero dc field, with an ac field of 10 Oe.

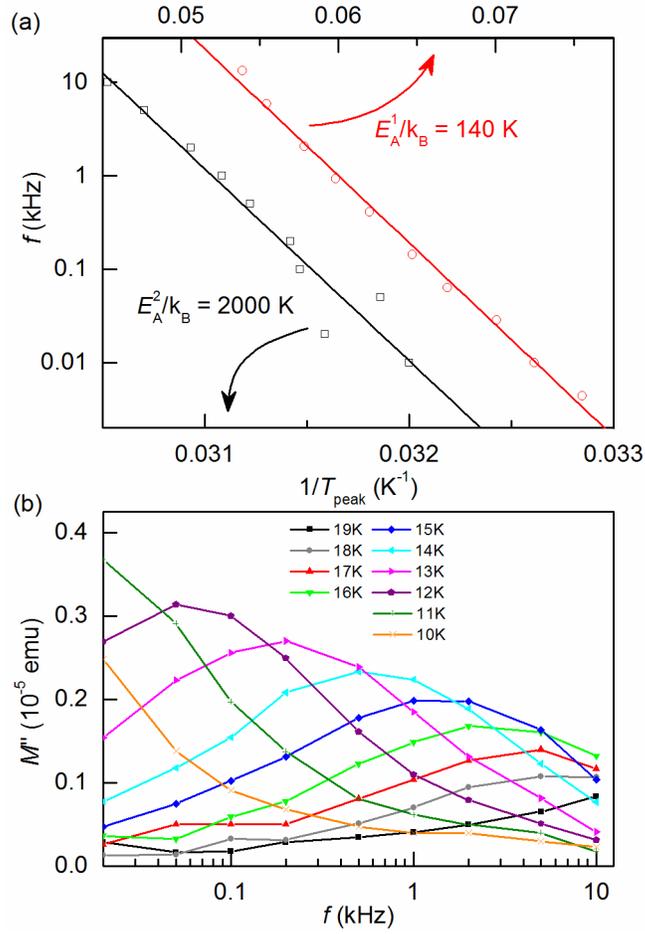

FIG. 8 (color online). (a) Fits as described in the text to the frequency dependence of the temperatures $T_{peak}$ at which the peaks in $M'$ are observed in the ac magnetization data of $SrCaCoOsO_6$. (b) Frequency dependence of $M''$ for the lower temperature transition.

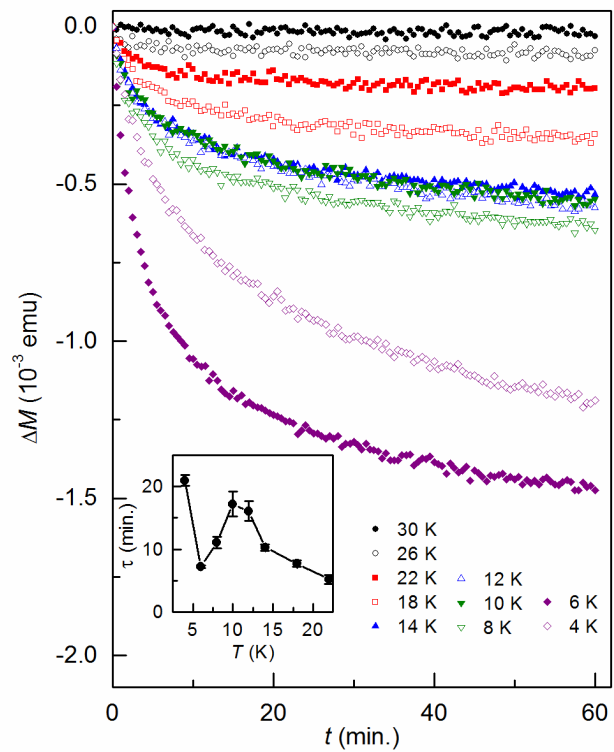

FIG. 9 (color online). Time dependence of the dc magnetic moment measured after cooling from 50 K in a magnetic field of 10 kOe and then reducing the field to zero. The inset contains the time constant determined from fits to the data.